\documentclass[10pt]{iopart}
\usepackage{graphicx}

\begin{document}
\title[Quasi-classical model of dynamic molecular structure]
{Quasi-classical model of dynamic molecular structure and non-destructive wavepacket manipulation by ultrashort laser pulses}

\author{W A Bryan$^{1}$, C R Calvert$^{2}$, R B King$^{2}$, G R A J Nemeth$^{1}$, J B Greenwood$^{2}$, I D Williams$^{2}$ and W R Newell$^{3}$}

\address{1) Department of Physics, Swansea University, Singleton
Park, Swansea SA2 8PP, UK}
\address{2) Department of Pure and Applied Physics, Queen's University Belfast,
Belfast BT7 1NN, UK}
\address{3) Department of Physics and Astronomy, University College
London, Gower Street, London WC1E 6BT, UK}

\ead{w.a.bryan@swansea.ac.uk}

\begin{abstract}
A quasi-classical model (QCM) of molecular dynamics in intense femtosecond laser fields has been developed, and applied to a study of the effect of an ultrashort `control' pulse on the vibrational motion of a deuterium molecular ion in its ground electronic state. A
nonadiabatic treatment accounts for the initial ionization-induced vibrational population caused by an ultrashort `pump' pulse. In the QCM, the nuclei move classically on the molecular potential as it is distorted by the laser-induced Stark shift and transition dipole. The nuclei then adjust to the modified potential, non-destructively shifting the vibrational population and relative phase. This shift has been studied as a function of control pulse parameters. Excellent agreement is observed with predictions of time-dependent quantum simulations, lending confidence to the validity of the model and permitting new observations to be made. The applicability of the QCM to more complex multi-potential energy surface molecules (where a quantum treatment is at best difficult) is discussed.
\end{abstract}

\section{Introduction}
A quantum wavepacket is a coherent superposition of states,
which can exhibit a time-dependent localisation of amplitude in one
or more degrees of freedom. In a significant advancement in understanding the dynamic structure of small molecules in intense laser fields (recently reviewed in \cite{post}), there has been significant recent
interest in creating a vibrational wavepacket in simple (few
electron) molecules. With the advent of few-cycle near infra-red
(NIR) laser pulses, a number of experimental groups have carried out studies to characterize vibrational wavepackets in hydrogenic molecular ions \cite{alnas, ergler1, ergler2, feuer2, mcken1, calv1, bryan2}. This type of system is particularly attractive as
a prototypical system for testing theoretical descriptions. Given
the simplicity of the hydrogenic molecular ions, it is relatively
straightforward to treat the time-dependent Schrodinger
equation in a field-free environment (within the Born-Oppenheimer and dipole approximations), which leads to the identification of quantum revivals of vibrational wavepackets. Revival occurs when all frequency components of a wavepacket
recover their initial phase conditions, so in the case of a
vibrational revival in hydrogenic molecular ions, wavepacket amplitude is relocalized at its initial location in the internuclear coordinate ($\it{R}$).

Experimental and theoretical manipulations of the wavepacket have
been reported, often with relatively weak ultrashort NIR
laser pulses. Niikura and co-workers have employed NIR laser pulses to perform experimental investigations of the laser induced dipole forces to influence vibrational motion \cite{niik3, niik2, niik1} in H$_2^+$ and D$_2^+$, focussing on measuring the ion fragmentation energy as the pump-control delay is varied. Thumm and co-workers have concentrated on finding an accurate quantum mechanical treatment of the behaviour of a vibrational wavepacket \cite{feuer1} and the influence of an ultrashort control pulse \cite{thumm, nieder2}. The authors in collaboration with Murphy and McCann have used a similar numerical treatment to investigate heating and cooling of vibrational populations \cite{murph3, murph2, murph1}, recently leading to the prediction of the `quantum chessboard' effect \cite{calv2}.

The demonstration of such a degree of control is
very attractive for a number of reasons. Firstly, manipulating the
phase and amplitude of a vibrational wavepacket could have
applications for a single molecule quantum computer \cite{molQC} whereby the qubit is stored in the superposition of vibrational states. While
this approach may not appear to be as technologically accessible
or feasible as qubits encoded in atomic systems coupled with photons \cite{atomQC}, the large number of
coupled vibrational and rotational modes may allow computations
over a large Hilbert Space. Furthermore, it is conceivable that
information be encoded not only into the amplitude of a vibration
or rotational state, but also the relative phase of the state
could be exploited.

Secondly, the creation and control of a vibrational wavepacket has
major implications for quantum structural dynamic control (observing and defining the electronic and nuclear configuration of molecules
with light). By creating a multi-dimensional vibrational wavepacket
with a `pump' laser pulse then directing the evolution of the
wavepacket, the nuclear co-ordinates of a molecule could
potentially be controlled. The electronic state of the system
could then be probed in a time-resolved manner with an XUV attosecond pulse, and the
nuclear-electronic coupling investigated beyond the Born-Oppenheimer approximation. Alternatively, once a wavepacket has been initiated, an additional photon field could
alter the electronic state and the influence on the nuclear motion
tracked.

As the wavepacket is a superposition of vibrational
states, the transfer of population between states defines how the
wavepacket evolves. By demonstrating population transfer with an
ultrashort pulse in a controlled manner, the molecule can be
vibrationally heated or cooled arbitrarily \cite{murph3, nieder2}. Additional control pulses could lead to an optically tailored population distribution. This leads
to applications in photochemistry. Ultimately, as wavepacket motion is defined by a potential energy surface (PES), if it is possible to temporally and spatially
resolve the wavepacket, the Fourier transformation will return the PES \cite{feuer2}. While feasible and relatively straightforward for a one-dimensional system, the possibility for
such a measurement in a complex molecule is extremely appealing,
as knowledge of the PES reveals the electronic configuration,
defining reactivity. This then opens possibilities for
dynamically controlling molecular structure by specifically
populating certain vibrational and electronic states. If the electronic state changes while the wavepacket oscillates, the PES will change nonadiabatically. Observation of wavepacket motion during this modification will allow the non-Born-Oppenheimer dynamics to be quantified.

\section{The quasi-classical model in one dimension}
The QCM is broken into a series of steps. The creation of a
coherent superposition of states is modelled by extending the
nonadiabatic ionization model of Yudin and Ivanov \cite{yudin} to a
molecular system, allowing the vibrational state population to be
evaluated. A non-interacting classical ensemble is then created,
with weighted quantized vibrational levels to reflect the initial
state of the molecule. The initial internuclear positions of the
elements of the ensemble are defined by the intersection of the
ground electronic state of the molecular ion and the energy
corresponding to the mid-point between successive vibrational states $\it{v}$ = 0, 1, 2 .... For example, the (eigen)energy of $\it{v}$ = 3 is -0.0782 au (where zero energy is defined by dissociation asymptote) and the (eigen)energy of $\it{v}$ = 4 is -0.0719 au. The initial position of element of the ensemble corresponding to $\it{v}$ = 3 is $\it R$ = 1.439 au, found by taking the mid-point of these two energies and finding the corresponding point on the PES. The wavepacket evolution is then approximated by allowing the
classical ensemble to propagate on the internuclear PES. Now, applying a `control' pulse to the propagating ensemble causes a time-dependent deformation of the PES \cite{niik3}: at small R this is function of the transition
dipole, and at large R is a function of the AC Stark shift. The
resulting deformation of the potential accelerates or decelerates
components of the ensemble depending on their direction of motion,
transferring energy into or out of the system. Moving away from the classical interpretation, this transfer of energy is directly analogous to a Raman process, however is continuous (rather than discrete) and applies throughout the duration of the control pulse. The influence of this energy transfer is two-fold, changing the relative phase of
the ensemble components with respect to the unperturbed system,
and transferring population between vibrational states. The transfer of population occurs when the motion of an ensemble component is sufficiently perturbed by the control pulse that it takes on the characteristics of a higher- or lower-lying vibrational state (i.e. amplitude, frequency). Each of these processes will be discussed in detail.

\subsection{Molecular ionization in a few-cycle pulse}
As with recent experiments, in our prototypical molecular ion the
vibrational wavepacket is created by the multiphoton or tunnel
ionization of the precursor neutral molecule by a few-cycle NIR
(800 nm) pump pulse. This effectively transfers vibrational
population from the ground electronic (1s$\sigma$) and vibrational
state (v' = 0) of the neutral (D$_2$) to a range of vibrational
states (v = 0, 1, 2 ...) in the ground state ion (1s$\sigma_g$).
The distribution of states depends on the probability of
ionization and the overlap between the ground state wavefunction
in the neutral and the available vibrational states in the ion.

Applying the Franck-Condon principle, we treat ionization as a
`vertical' transition, thus the energy difference between the
neutral and ionic potentials approximates an $\it R$-dependent
ionization potential, $\it I_P(R)$ \cite{saenz}. The nonadiabatic
ionization model of Yudin and Ivanov \cite{yudin} is employed to predict
P$_{ion}$(R), the ionization probability as a function of $\it
I_P(R)$ by calculating the ionization rate, $\Gamma(\it{t})$. Here, the time-step in $\it{t}$ = 0.05 fs. Ionization as a
function of internuclear separation requires non-zero population
in the neutral ground state to be promoted to the ion, thus the
probability of population transfer is represented by:
\begin{equation}
P_{trans}(R) = P_{ion}(R) \times |\psi(R)_{v' = 0}|^2
\end{equation}
where $\psi(\it{R})_{v' = 0}$ is the neutral molecule ground state
wavefunction amplitude as a function of internuclear separation
$\it{R}$. This is approximated by a Gaussian distribution.

While this may not be the an exact description of molecular
ionization, it is a useful treatment as it is a reasonable
approximation to the underlying physics, is mathematically
straightforward, is applicable to both the multiphoton and tunnel
ionization regimes, and holds for ultrashort laser pulses. This is
in contrast to the ADK model \cite{adk}, which is commonly used in this context but only applies in the tunnelling regime for long laser pulses. As discussed later, the
QCM can be extended to complex systems on the condition that the
PES of the neutral and ionic states are known along the
co-ordinate frames of interest, such that the ionization potential
can be calculated at all points. Furthermore, ionization by an
attosecond UV / XUV pulse can also be included, as long as the
active electronic transitions are identified across the full
bandwidth of the pump pulse.

Returning to a NIR pump, the population of vibrational states in
the molecular ion is predicted by projecting the probability of
population transfer $\it{P}_{trans}(\it{R})$ onto the available
vibrational levels in the molecular ion over all internuclear
separations. The outcome of this prediction is presented in figure
1: several relative population distributions are presented for a 7
femtosecond NIR few-cycle pump pulse as the pump intensity is
varied between 2$\times$10$^{13}$ Wcm$^{-2}$ and
5$\times$10$^{14}$ Wcm$^{-2}$. There is a considerable variation
of the distribution of vibrational probability with intensity;
comparable theoretical findings have been reported recently
\cite{mads, urb}. Only the lowest vibrational states are populated at low
intensity, and as the intensity increases, the population
distribution tends to the Franck-Condon distribution (thin line).
This is the result of the variation of ionization rate with
internuclear separation and intensity.

\begin{figure}
\includegraphics[width=320pt]{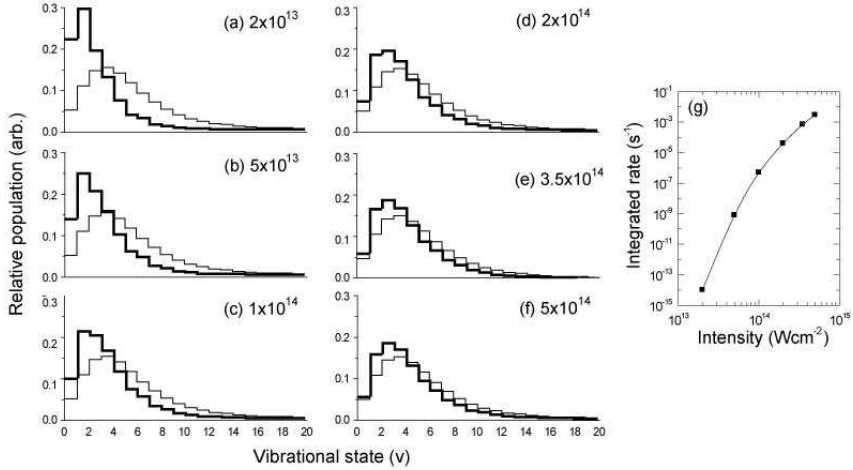}\\
\caption{Relative vibrational state populations as the intensity
of a 7 fs pump pulse is increased from (a) 2$\times$10$^{13}$
Wcm$^{-2}$ to (f) 5$\times$10$^{14}$ Wcm$^{-2}$. The Franck-Condon
distribution of states is indicated by the thin line. The largest
discrepancy between Franck-Condon and the predicted distributions
occurs at low pump intensity. The variation of ionization rate integrated over the pump pulse duration is shown in (g), an indication of the total number of ions generated at a particular intensity. The square markers indicate the intensities (a) to (f).}
\end{figure}

While there is a significant variation in the relative population
in the molecular ion, the overall ion production rate varies even
more significantly as demonstrated in figure 1(g). Integrating the ionization rate over the
duration of the pump pulse gives an idea as to the relative number
of ions generated at the peak of $|\psi(\it{R})_{v' = 0}|$ (i.e.
the maximum wavefunction amplitude in the ground electronic and
lowest vibrational state of the neutral precursor).
Therefore as the pump intensity is increased, there is a shift in
vibrational population to the higher vibrational states and the
overall number of ions generated increases dramatically. However this process
has an upper limit, meaning it is almost impossible to generate a
significant number of molecular ions with a perfect Franck-Condon
distribution of vibrational states. Even with a 7 fs duration
pump, at intensities greater than 2$\times$10$^{14}$ Wcm$^{-2}$,
the pump pulse can populate then immediately dissociate the
molecular ion.

\subsection{Propagation of ensemble components on the PES}
The QCM treats the motion of each element of the ensemble as a Newtonian particle accelerating on a PES, whereby the acceleration of a unit mass at internuclear position $\it{R}$ is
related to the PES $\it{V}(\it{R}$) by:
\begin{equation}
\frac{d^2 R}{dt^2} \propto \frac{d V(R)}{dR}
\end{equation}
where $\it{t}$ is time. The presented discussion is confined to
one dimension to be applicable to the hydrogenic molecular ion,
however equation (2) holds for any number of degrees of freedom.
Indeed, in complex polyatomics, classical ensemble simulations are
employed when unravelling the femtochemistry.

It is assumed that the PES is smooth and well-behaved, such that
the numerical differential of a relatively coarse sample of the
PES is an accurate representation of the actual continuous
differential. In the one dimensional case, a five point stencil is employed for the numerical differentiation. The key
requirement is that the discrete numeric differential can be
reliably and efficiently interpolated; in this work a natural
Spline function is employed, however it is suggested that the PES
be sampled to a sufficient spatial resolution that the selection
of the interpolation method is essentially arbitrary on the
condition that a smooth continuous output results.

Now, by discretely calculating the equations of motion of the unit
mass as influenced by the differential of the PES, the ensemble
propagation is simulated as a function of time. It is assumed that
the unit mass is initially stationary following projection onto
the ionic PES, which is reasonable considering that the tunnel
time is much shorter than the characteristic time-scale for
wavepacket propagation.

\begin{figure}
\includegraphics[width=252pt]{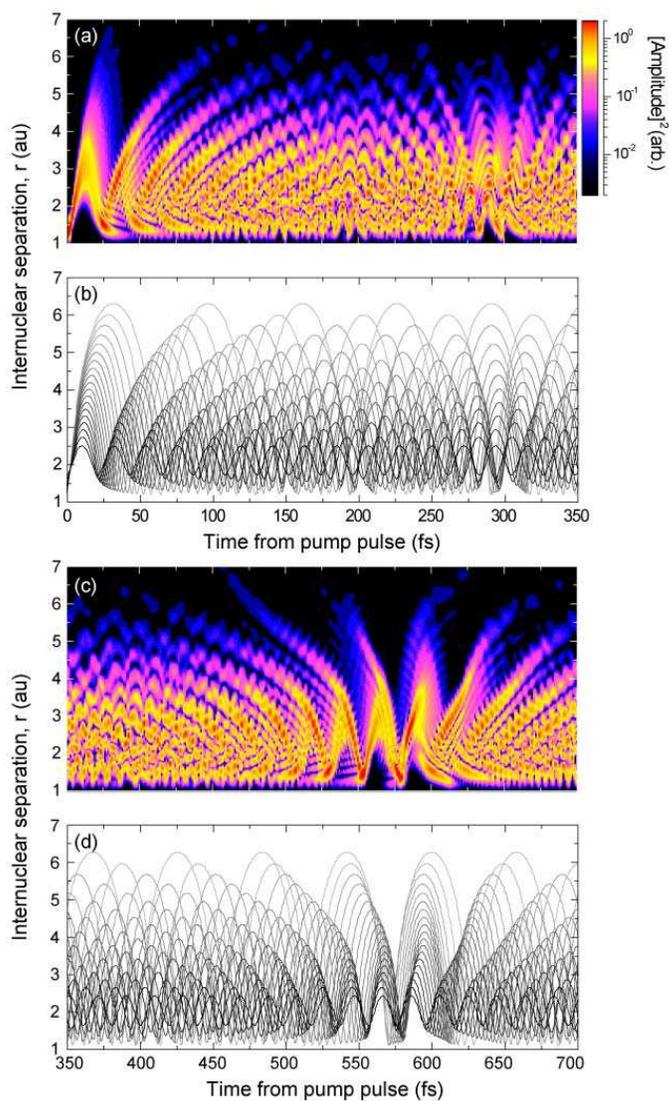}\\
\caption{Comparison between the time-dependent solution of
the Schrodinger equation for the evolution of the vibrational
wavepacket (v = 0 to 18) in D$_2 ^+$ (a, c, colour panels) and the
QCM model (b, d, trajectories). The initial dephasing (a and b)
and revival (c and d) of the vibrational wavepacket is shown. The
QCM reproduces the features of the TDSE solution, as there is a
bunching of the trajectories in regions where the solution of the
TDSE is localized. The false colour panels are on a logarithmic
scale, and the grey-scale of the trajectories has been adjusted to
illustrate similarities.}
\end{figure}

The motion of the ensemble (and hence the approximation to the
wavepacket motion) is then predicted by allowing the unit masses to propagate as a function of time. The result of propagating the quasi-classical
model for a large number of populated vibrational states is
presented in figure 2, and a visual comparison made to the solution of the time-dependent Schrodinger equation (TDSE) for D$_2^+$ within the Born-Oppenheimer and dipole approximations, as published in \cite{bryan2, mcken1, calv1}. Clearly as the QCM generates a series of trajectories, and the quantum treatment results in a continuous wavefunction (both of which evolve in time) a direct numerical comparison is difficult. However, from figure 2 it is apparent that the QCM captures the essence of the underlying motion.

Figure 2 presents two subsets of the now familiar wavepacket dephasing
(figure 2a) and revival (figure 2b). Comparing the QCM
trajectories and the predicted TDSE-derived wavepacket amplitude
as a function of internuclear separation, a remarkable agreement
is found, whereby intense features in the false colour
representation of the wavepacket are apparent in the overlapping
or closely-spaced QCM trajectories. The ensemble representation of
a wavepacket is therefore reasonable, as is the numerical
treatment of the ensemble motion.

The constant maximum and minimum amplitudes
of oscillation of individual ensemble elements over a number of
periods indicates the numerical robustness of the QCM. Indeed,
this is a method to efficiently define the temporal resolution, $\Delta \it{t}$ of the calculation. If $\Delta \it{t}$ is too
small, the calculation will take a prohibitively long time; too
large and the magnitude of oscillation will vary, indicative of the cumulative effect of inaccurately calculating the derivative of the PES. A value of
$\Delta \it{t}$ = 0.1 fs was found to be a good compromise.

\begin{figure}
\includegraphics[width=370pt]{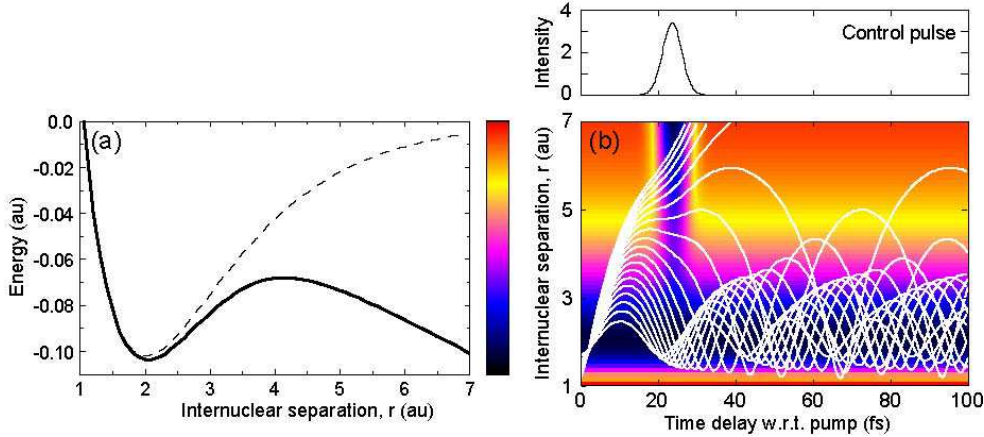}\\
\caption{Potential energy surface distortion by the control pulse
and associated ensemble perturbation. (a) With a control pulse
intensity of $\it{I_{control}}$ = 3.5 $\times$10$^{13}$
Wcm$^{-2}$, the 1s$\sigma_g$ potential is heavily distorted (solid line) as compared to the field-free situation (dashed line). (b) The
distortion of the PES as a function of time when a 7 fs control pulse
is applied at 24 fs on a colour scale, referenced to the energy
scale of (a). The trajectories predicted by the QCM are overlaid
on the potential; the distortion of all trajectories is heavily
dependent on state amplitude and phase relative to the control
pulse (shown in units of 10$^{13}$ Wcm$^{-2}$). The maximum in the PES induced by the control pulse around 4 au present in (a) becomes a saddle point as a function of time in (b). }
\end{figure}

\subsection{Quantifying deformation of the PES}
For the laser intensities considered here, it is assumed that the structural dynamics of the hydrogenic molecular ion can be accurately predicted by wavepacket motion on,
and coupling between, the two lowest lying potential energy
surfaces (bound state 1s$\sigma_g$ and dissociative state
2p$\sigma_u$). This approach is not unique to the QCM, but is
accepted because the 1s$\sigma_g$ and 2p$\sigma_u$ states are
energetically isolated from the electronically excited states of
the ion. Clearly this will be a contentious issue when dealing
with complex many-electron molecules, as a large number of
overlapping bound and dissociative states exist; this point is addressed later.

We quantify the distortion of the 1s$\sigma_g$ state \cite{die} as:
\begin{equation}
V(R,t) = \frac{V_{2p\sigma_u}(R) - V_{1s\sigma_g}(R)}{2} +
\sqrt{\frac{\Delta V(R)^2}{4}+ \Omega^2}
\end{equation}
where $\Omega = \it{R}|\it{E}(\it{t})|$/2, and we define
$\it{E}(\it{t})$ as the time-dependent electric field induced by
the control pulse, hence is a function of
$\it{I_{control}}(\it{t})$. It is important that $\it{E}(\it{t})$
be recovered from a FROG \cite{frog} or SPIDER \cite{spider} measurement,
allowing a realistic quantification of the temporal intensity and
phase and/or the spectrum and spectral phase.

In figure 3(a), the field-free PES and a typical laser-distorted
PES are shown. If the potential is too
dramatically distorted, the ensemble will no longer be bound by a
concave PES, and dissociation will occur (as seen in the case of
the five highest vibrational states in Fig. 3b). However, the
dynamic nature of the distortion should be considered: if the
control pulse is shorter than approximately a quarter the period
of a particular vibrational beat frequency, that component of the
ensemble will be resilient to dissociation if it is in the
vicinity of the inner turning point.

The distortion of the PES by the control pulse causes a
time-dependent variation of $\it{dV}(\it{R})$/$\it{dR}$, causing
the ensemble components to experience an additional acceleration,
the direction and magnitude of which depends on the direction the
component is moving and its location on the PES. This is
illustrated in figure 3(b), whereby a 7 fs FWHM Gaussian control
pulse of peak intensity $\it{I_{control}}$ = 3.5$\times$10$^{13}$
Wcm$^{-2}$ is applied 24 fs after the wavepacket is created by the
pump pulse (i.e. $\it{t}$ = 0). The control pulse couples the
V$_{1s\sigma_g}$(R) and V$_{2p\sigma_u}$(R) PESs, inducing an AC
Stark shift, resulting from the gradient of the laser-induced
electric field resolved along the internuclear bond. If the
control pulse intensity $\it{I_{control}}$ $<$ 10$^{14}$
Wcm$^{-2}$, it is possible for the molecular ion to survive the
control pulse (i.e. not undergo dissociation), and the distortion
of the potential energy surfaces will transfer energy to or from
the ensemble.

As the control pulse is applied, the vibrational trajectories evolve on the laser-distorted PES as a function of internuclear separation, intensity and time as evidenced in figure 3(b). Even a moderate intensity (as compared to
intensities required to cause significant atomic ionization)
causes a significant distortion of the PES; such a distortion is
dynamically modelled by numerically propagating the ensemble
throughout the control pulse. Importantly, wavepacket dynamics
initiated by a totally arbitrary control pulse or pulses can be
quantified, as the laser-distorted PES is calculated at each time
step. For comparison, the ensemble propagation on the undistorted
potential has already been given in figure 2. It can be seem that
the phase and amplitude of the ensemble components are shifted by
the control pulse.

\subsection{Variation of vibrational state phase and population}
In the QCM, with no control pulse present, the natural
motion of the classical ensemble would result in a wavepacket that
would continuously dephase and rephase, as is predicted and well
understood from the solution of the Schrodinger equation. The
action of the control pulse is to artificially perturb the
distribution of vibrational population and the phase of each
ensemble component with respect to the unperturbed motion.

\begin{figure}
\includegraphics[width=370pt]{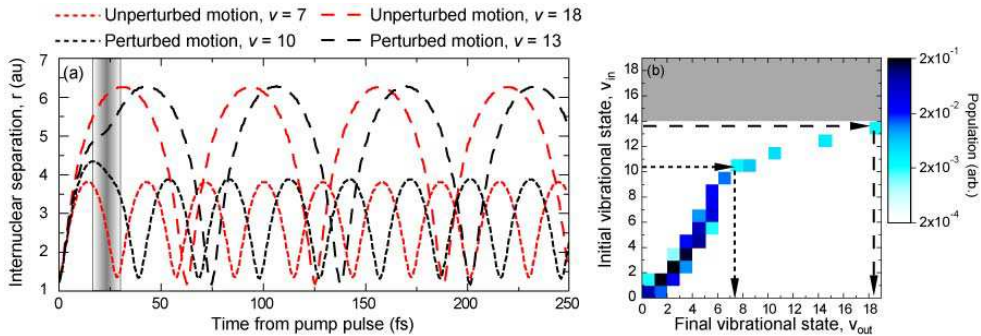}\\
\caption{Vibrational population transfer in QCM. (a) The
application of a 7 fs duration control pulse with an intensity of
3.5$\times$10$^{13}$ Wcm$^{-2}$ control pulse at 24 fs causes both
positive and negative population transfer. The initial state v =
10 is down-shifted into resonance with the v = 7 state by the
control pulse (red and black short dashed lines), while the v = 13
state is up-shifted into resonance with the v = 18 state (red and
black long dashed lines). The shaded region indicates the intensity of the control pulse and the vertical lines indicate when the intensity is 3.5$\times$10$^{12}$ Wcm$^{-2}$. (b) The vibrational population matrix
conveniently illustrates the action of the control pulse. Each
point on the matrix indicates the likelihood of population
transfer on a logarithmic colour scale, and the grey region for v
$>$ 14 indicates states are no longer bound. Projecting the sum of
the matrix vertically returns the final vibrational population.}
\end{figure}

As is illustrated in figure 4, the distortion of the
molecular PES by the control pulse causes the QCM trajectories to
vary. If additional energy is transported into a trajectory, the
amplitude of the oscillation increases, along with the vibrational
period, hence the trajectory takes on the characteristics of a
higher lying state. We propose that this increase in energy is equivalent
to population being transferred from the v = 13 to v = 18 state.
Similarly, the control pulse can simultaneously remove energy from
the system, causing a decrease in vibrational state, as seen by
the transfer of the v = 10 to v = 7 state.

There is a clear phase difference between the vibrational motions
highlighted in figure 4(a) following the control pulse. Depending
on the intensity and arrival time of the control, this phase
difference can be positive or negative. By the up- and
down-shifting of vibrational population, it is possible that
multiple ensemble components will end up in a particular
vibrational state, and the phase and relative population of each
state must be considered. Variations in the phase and population
are quantified by making a numerical comparison between the
perturbed and unperturbed trajectories. Importantly, this comparison must be made once the intensity of the control pulse has dropped to essentially zero, such that the trajectories are only defined by the static PES.

Using the unperturbed trajectories as a reference (figure 2), the square of
the difference between unperturbed and perturbed motions is
calculated, defining a quality of fit parameter. An arbitrary time offset is introduced and varied over a range greater than the period of the highest vibrational state. The minimum values of the fitting parameter
therefore returns the distribution of final vibrational states and
time offset that best represent the ensemble after the action of
the control pulse. The time offset is directly related to the
phase via the known periods of the unperturbed motions.

The shifting of population is not treated in a totally discrete
manner, rather if a final state is equally represented by two
closely spaced states, the initial population will be shared
according to the ratio of the fitting parameters. To determine the
final distribution of vibrational states, the populations in each
state are summed, as illustrated in the vibrational population matrix shown in figure 4(b). The redistributed initial population (caused by a pump intensity of
10$^{14}$ Wcm$^{-2}$) by the control pulse is apparent. The vibrational population matrix is a convenient visualization of the influence of the control pulse. With no control pulse present, the vibrational population matrix would be a diagonal line as $\it{v}_{in}$ = $\it{v}_{out}$.

On transferring between states, we place no constraint on the
relative phase between the unperturbed and perturbed motions of
the trajectories. So, the resultant phase with respect to the
natural motion can be varied depending on when a vibrational state
undergoes state transfer, hence dependent on the intensity and
delay of the control pulse. If a number of ensemble components
contribute, a weighted mean is calculated, depending on the
relative initial population.

\section{Comparison with established results}
\begin{figure}
\includegraphics[width=370pt]{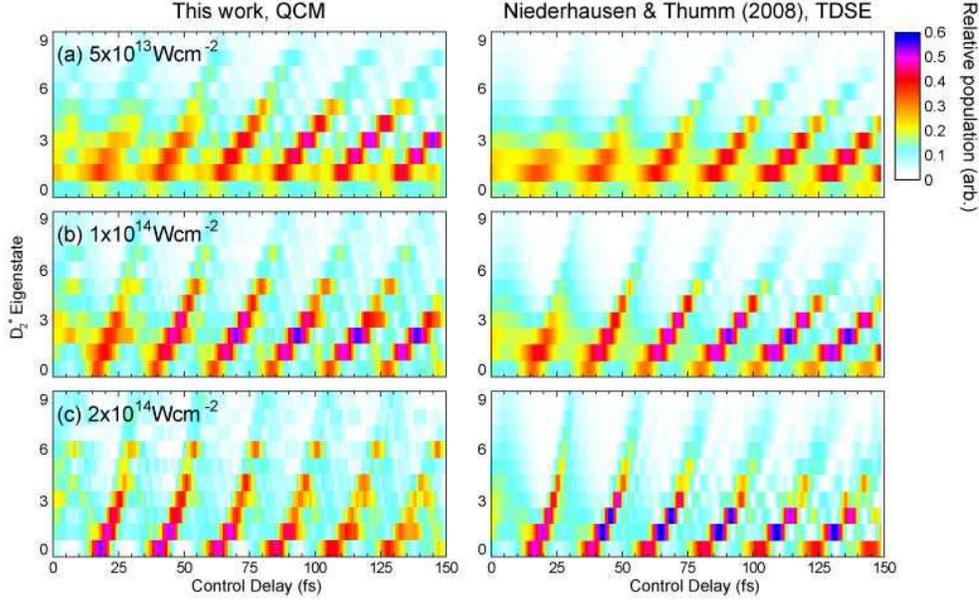}\\
\caption{Final vibrational state population distributions for three different
control pulse intensities as the temporal separation between the pump and control pulse is varied. The results of Niederhausen and Thumm, Phys. Rev. A $\bf{77}$ , 013407 (2008) are reproduced for comparison with the QCM, which reproduces the periodicity and
relative shift of vibrational population as the delay between the
pump (6 fs, 1$\times$10$^{14}$ Wcm$^{-2}$) and the control is
varied.}
\end{figure}
While figure 2 has demonstrated that the QCM can successfully reproduce the unperturbed motion of a vibrational wavepacket by propagating the ensemble, it is vital to quantify the ability of the QCM to accurately describe the action of the control pulse. To this end, we compare the output of the QCM to established theoretical results.

In figure 5, systematic scans of the control delay and intensity
with respect to the pump pulse are presented for direct comparison
with the reproduced results of Niederhausen and Thumm, Figure 5, Phys. Rev. A $\bf{77}$, 013407
(2008) \cite{nieder2}. In both cases, the control pulse is 6 fs in duration and the initial vibrational state distribution is calculated using tunnelling theory. In \cite{nieder2}, the TDSE was solved within the Born-Oppenheimer approximation, and the control pulse causes Raman transitions between vibrational states. A good agreement is found (especially considering how disparate the two numerical techniques are), indicating that the QCM accurately captures the modification of the vibrational population. Approximating the evolution of a wavepacket under the influence of an ultrafast control pulse as a quasi-classical ensemble is therefore justified.

\section{QCM prediction of vibrational population and phase}
In figure 6 the vibrational population matrices, initial and final
vibrational state populations and final phases are shown as the
intensity of the control pulse is varied. The initial vibrational state corresponds to a pump pulse intensity of 1$\times$10$^{14}$ Wcm$^{-2}$, and the control pulse
arrives at 28 fs. At the lowest intensity of 5$\times$10$^{12}$
Wcm$^{-2}$, the near diagonal distribution of vibrational
population indicates the initial and final states are similar, as
the control pulse perturbation is minimal. Nonetheless the phase
of the ensemble is altered, most significantly for the highest
lying states. At the highest control intensity of
1$\times$10$^{14}$ Wcm$^{-2}$, the matrix indicates a dramatic
redistribution of population, skewing the maximum in the
population from $\it v$ = 1 to $\it v$ = 6. As is apparent from the matrix,
initial states with $\it v$ $>$ 12 are dissociated by the relatively
intense control pulse.

\begin{figure}
\includegraphics[width=345pt]{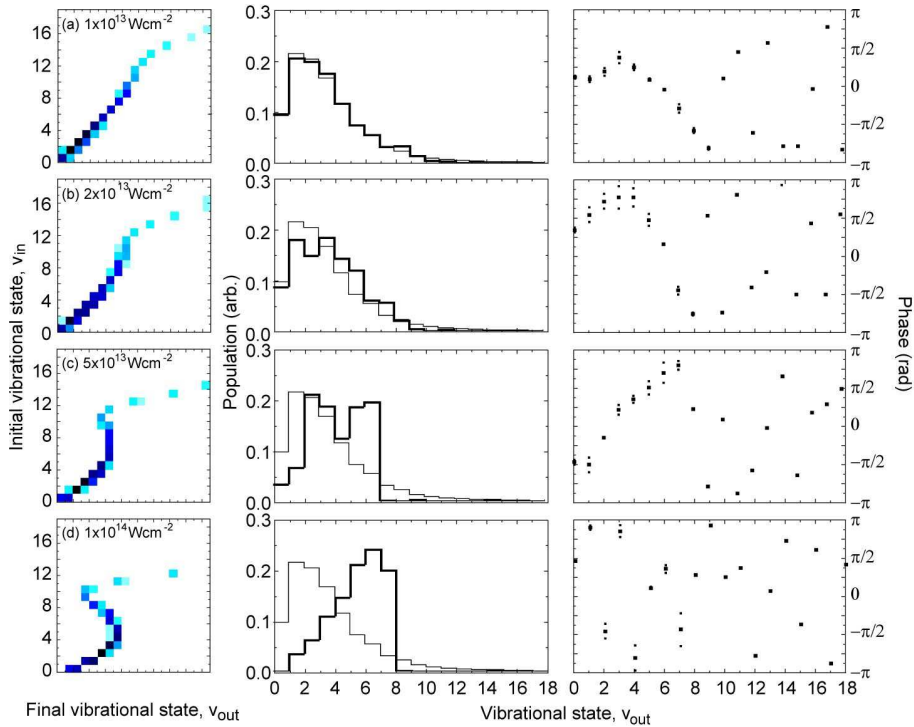}\\
\caption{Population transfer matrices, vibrational populations and phases as
the intensity of the control pulse is varied. The control arrives
28 fs after the ensemble is generated, and is 7 fs in duration. The vibrational population generated by the 1 $\times$ 10$^{14}$ Wcm$^{-2}$ pump pulse is indicated by the thin line. The weighted mean is indicated by the large markers, and the small markers indicate 1 standard deviation.}
\end{figure}

\begin{figure}
\includegraphics[width=345pt]{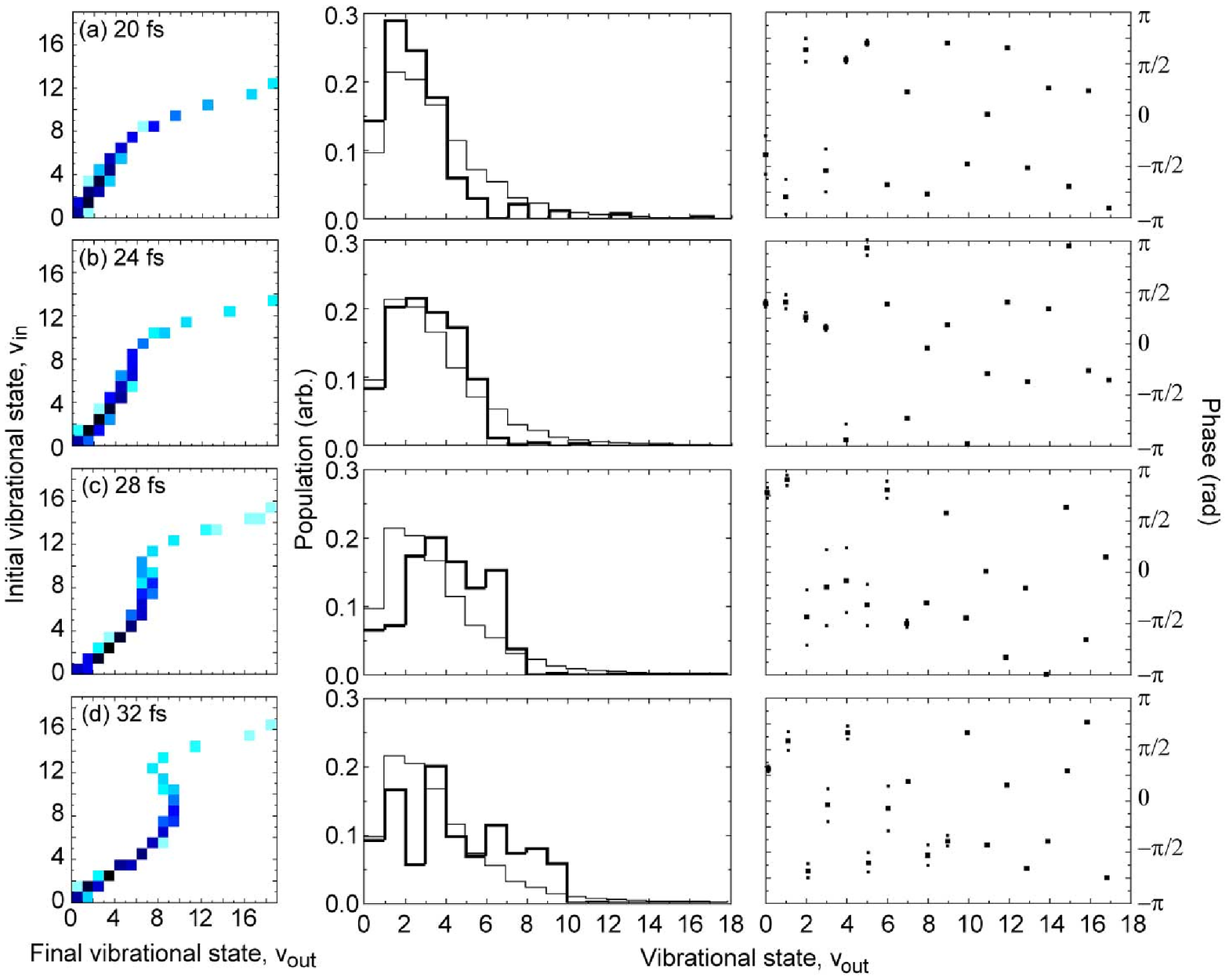}\\
\caption{Population transfer matrices, populations and phases as
the arrival time of the control pulse is varied. The control pulse
has an intensity of 3.5 $\times$ 10$^{13}$ Wcm$^{-2}$, and is 7 fs
in duration. The vibrational population generated by the 1 $\times$ 10$^{14}$ Wcm$^{-2}$ pump pulse is indicated by the thin line. The weighted mean is indicated by the large markers, and the small markers indicate 1 standard deviation.}
\end{figure}

Figure 7 shows how the vibrational population matrices, initial
and final vibrational state populations and final phases vary as
the delay between the pump (1$\times$10$^{14}$
Wcm$^{-2}$) and control (3.5$\times$10$^{13}$ Wcm$^{-2}$) pulses is varied in 4 fs
steps. Even with a fixed control
intensity, the degree by which the final vibrational population
and phase can be varied is dramatic. Changing the control arrival
time from 28 to 32 fs induces a significant shift, which we suggest
should be experimentally observable.

The result of applying a control pulse can be resolved by imaging
the wavepacket (i.e. by destroying the molecular bond). As with
previous studies, the temporal evolution of the wavepacket is
revealed by an intense probe pulse. As discussed earlier, an
intensity of greater than 2$\times$10$^{14}$ Wcm$^{-2}$ would be
optimal to access the majority of bound states. Varying the delay
between the pump and control changes the distribution of states in
the superposition. Varying the delay between the control and
probe, then Fourier transforming the resulting time-dependent
fragmentation yield will reveal the influence of the control
pulse.

The final stage of the QCM is then to predict the fragmentation
yield. This can be done by modelling the distortion of the PES by
the probe pulse, then identifying the vibrational states that do
not survive the probe. If the probe intensity is high enough this
will be almost all states. However, this calculation is time
consuming: a far more efficient method has been recently
demonstrated \cite{calv2}. The dissociation yield can be accurately
estimated for a chosen pulse intensity and duration, by identifying the elements of the ensemble that are above some critical internuclear separation during the probe pulse.

\section{Applicability of the QCM to complex molecules}
It is feasible to extend the QCM to multi-electron systems pumped
with an attosecond UV - XUV pulse. A coherent attosecond pulse can
only be supported by an ultrabroadband photon field; as a result a
wide range of electronic and vibrational states will be populated
on absorption of the pump pulse. On the condition that the PESs of
all populated electronic states is known, the ensemble and hence
the wavepacket can be modelled. The application of a control pulse
will distort the PESs; this distortion should be on a comparable time-scale to the characteristic vibrational motion, therefore it suggested that the control pulse be NIR femtosecond radiation rather than XUV attosecond in nature. To correctly describe how the wavepacket
evolves, the coupling between bound and dissociative states must
be addressed, and is nontrivial. This situation is further
complicated if the molecule is polyatomic, as this coupling is
dependent on the angle between the various bonds and the
polarization direction of the control pulse.

The application of the QCM to a polyatomic molecule is currently
underway, and is somewhat more involved than the current work.
Firstly, the electronic and vibrational states populated by the
probe pulse must be identified. This requires the calculation of
accurate PESs for all degrees of freedom; such calculations are
possible with modern ab initio quantum chemistry software \cite{GAMESS}.
Resonances between active states before and after the pump within
the bandwidth of the pump will then be considered in terms of
selection rules. It then is suggested that the QCM be applied to
those degrees of freedom with the deepest PESs. Naturally such a
calculation will be multidimensional, and the efficiency of which
should be judged against experimental results. The first test of
the application of the QCM to a polyatomic molecule will be the
observation and prediction of an unperturbed wavepacket. The
influence of the control pulse should then be quantified by
resolving the applied electric field components along the active
co-ordinates.

The large number of vibrational and electronic states that could
potentially be populated by a coherent attosecond pulse appears to
be a problem for experimentally identifying the populated states.
The emerging study of quantum structural dynamics (QSD), the
time-dependent observation and control of nuclear and electronic
motions, requires such identifications to be made. We propose an
interesting solution, or at least a viable approach: it is
possible to exploit the varying coupling between bound and
dissociative states under the influence of the control pulse to
carry out exactly the state identification required by QSD
investigations. It is highly unlikely that the energy difference
between the available bound and dissociative electronic states
would be identical over all internuclear separations. Therefore,
by applying a control pulse of a known intensity, the different
degree of PES distortion will vary the amount of up- and
down-shift of elements of the ensemble depending on the electronic state. By then Fourier
transforming the probe delay dependent fragmentation yield, the
electronic states are potentially identifiable.

\section{Conclusion}
A quasi-classical model has been proposed that allows the quantification of wavepacket dynamics modified by an ultrashort laser pulse.
We have discussed how the vibrational phase and population are
adjusted by the control pulse, and a comparison has been made to established theoretical predictions. Systematic predictions of
wavepacket dynamics as the pump intensity and control delay and
intensity have been presented. Such results will be of interest to
groups attempting to experimentally detect the manipulation of a
wavepacket. Finally, application of the quasi-classical model to
attosecond studies of quantum structural dynamics in complex many-electron molecules is discussed.

\ack{We thank Thomas Niederhausen for communicating the data used in figure 5.
 This work was supported by the Engineering and Physical Sciences Research Council (EPSRC) and the Science and Technology Facilities Council (STFC), UK. CRC and RBK acknowledge financial support from the Department of Education and Learning, NI, GRAJN acknowledges financial support from EPSRC and STFC.}

\end{document}